\documentclass[10pt, a4paper]{article}
\usepackage[english]{babel}
\usepackage{graphicx, pdflscape, float}
\usepackage{amssymb}
\usepackage{xcolor}
\usepackage{setspace}
\usepackage[margin=0.75in]{geometry}
\usepackage{amsmath}
\usepackage{amsthm}
\newtheorem*{remark}{Remark}

\providecommand{\keywords}[1]
{\textbf{\textit{Keywords:}}#1}
\begin{document}
\doublespacing
\title{Estimation of the Parameters of Symmetric Stable ARMA and ARMA-GARCH Models}
\author{Aastha M. Sathe and N. S. Upadhye\\
Department of Mathematics, Indian Institute of Technology Madras,\\
Chennai-600036, INDIA\\
Email address: aasthamadonnasathe@gmail.com, neelesh@iitm.ac.in}

\date{\today}
\maketitle
\begin{abstract}
\noindent
In this article, we first propose the modified Hannan-Rissanen Method for estimating the parameters of autoregressive moving average (ARMA) process with symmetric stable noise and  symmetric stable generalized autoregressive conditional heteroskedastic (GARCH) noise. Next, we propose the modified empirical characteristic function method for the estimation of GARCH parameters with symmetric stable noise. Further, we show the efficiency, accuracy and simplicity of our methods through Monte-Carlo simulation. Finally, we apply our proposed methods to model the financial data.

\noindent
\keywords{ ARMA-GARCH models, stable distributions, parameter estimation, simulation, application }
\end{abstract}
\section{Introduction}
\label{sec1}
In finance and econometrics, important information about past market movements is modelled through the conditional distribution of return data series using the autoregressive moving average (ARMA) models.  However, in these models, the conditional distribution is assumed to be homoskedastic which poses challenges in the modelling and analysis of returns with time-varying volatility and volatility clustering, a commonly observed phenomena. This led to the development of  autoregressive conditional heteroskedastic (ARCH) and generalized ARCH (GARCH) models introduced by Engle \cite{eng} and Bollerslev \cite{bol} respectively. In empirical finance, the most important and widely used model is the combination of ARMA with GARCH, referred to as an ARMA-GARCH model (see \cite{zang}, \cite{zou}).

\noindent
In general, the noise term of  ARMA, GARCH and ARMA-GARCH models is assumed to be normal or Student-$t$ distribution, where the normal distribution is known for the desirable property of stability but fails to capture the heavy-tailedness of the data. On the other hand, the Student-$t$ distribution allows for heavier tails, but, when compared with the normal, lacks the desirable property of stability. These flaws, therefore, motivate us to use the family of stable distributions as a model for unconditional, conditional homoskedastic and conditional heteroskedastic return series distribution. Some of the significant and attractive features of stable distributions, apart from stability are heavy-tailedness, leptokurtic shape,  domains of attraction and skewness. For more details on stable distributions, see \cite{sam}. Hence, there is a need to explore the behaviour of ARMA, GARCH and ARMA-GARCH models with stable noise for effective modelling of the return data series which also involves estimation of the parameters of these models, and only a handful of estimation techniques based on stable noise are available. For details on the different estimation techniques used for such models, see \cite{mik}. Therefore, we feel that this article is an important contribution to the literature available on these models.
 
\noindent
In this article, we develop a method for estimating the parameters of ARMA model with symmetric stable noise and symmetric stable GARCH noise by modifying the Hannan-Rissanen Method \cite{han}. For the estimation of the GARCH parameters, we propose the modified empirical characteristic function method which is based on the method discussed in \cite{om}. The efficiency and effectiveness of the two proposed methods is validated through Monte Carlo simulation. For comparative analysis, we compare the modified Hannan-Rissanen method with the first two of the three specifc M-estimators of the parameters of ARMA models with symmetric stable noise introduced by Cadler and Davis \cite{cad}, namely, the least absolute deviation (LAD) estimator; the least squares (LS) estimator, and the maximum likelihood (ML) estimator. 
%
Amongst the three M-estimators, LAD is an excellent choice for estimating ARMA parameters with stable noise due to its robustness, computational simplicity and good asymptotic properties.  
Our method works at par with the LAD estimator both in terms of accuracy and computational simplicity. 

\noindent
The paper is organised as follows. Section \ref{sec2} gives a brief introduction to the stable distributions  along with the necessary definitions and notations. Section \ref{sec3} discusses the two new methods proposed for the estimation of ARMA, GARCH  and ARMA-GARCH parameters. Section \ref{sec4} deals with simulations and comparative analysis of the proposed methods. Section \ref{sec5} discusses application of our proposed methods to the financial data. Finally, Section \ref{sec6} gives some concluding remarks on the proposed methods.

\section{Preliminaries and Notations}\label{sec2}
\textbf{Stable distributions.} These distributions form a rich class of heavy-tailed distributions, introduced by Paul L\'evy \cite{lev}, in his study on the Generalized Central Limit Theorem. Each distribution, in this class, is characterized by four parameters, namely $\alpha$,  $\beta$,  $\sigma$, and $\delta$, which, respectively, denote the index of stability, skewness, scale and shift of the distribution. Their respective ranges are given by  $\alpha \in (0,2]$, $\beta \in [-1, 1]$, $\sigma > 0$ and $\delta \in \mathbb{R}$. For more details, see 
\cite{hybrid}. In this paper, we deal with symmetric stable distributions. We say that the distribution is symmetric  around zero if and only if $\beta=\delta=0$ and denote by $S\alpha S$. The characteristic function (cf) representation is given by
\begin{eqnarray}
\phi(t)&=& \begin{cases}
\exp\left\{-(\sigma|t|)^\alpha\right\},&
\alpha \neq 1,\\
\exp\left\{-\sigma|t|\right\}, & \alpha = 1.
\end{cases} \label{2par}
\end{eqnarray}
For simulations, we assume $\sigma=1$ for $S\alpha S$ distribution throughout.
\newpage

\noindent
Next, we discuss some important definitions and notations required for parameter estimation of ARMA, GARCH and ARMA-GARCH models. Let $\boldsymbol{\phi}=(\phi_1,\cdots,\phi_p)$ and $\boldsymbol{\theta}=(\theta_1,\cdots,\theta_q)$ denote the parameter vectors used in the ARMA model. For the GARCH model, the parameter  vectors used are  $\boldsymbol{a}=(a_1,\cdots,a_q)$, $\boldsymbol{b}=(b_1,\cdots,b_p)$, $\boldsymbol{a}_G=(a_1,\cdots,a_{q_G})$ and $\boldsymbol{b}_G=(b_1,\cdots,b_{p_G})$.

\noindent
Let the time series $\lbrace X_{t}\rbrace$ be an  \textbf{ARMA($p,q$) process with $S\alpha S$ noise}, denoted by $S\alpha S$-ARMA($p,q$), if 
\begin{equation}
X_{t}=\sum_{i=1}^{p}\phi_{i}X_{t-i}+\sum_{j=1}^{q}\theta_{j}\epsilon_{t-j}+\epsilon_{t}, \label{10}
\end{equation}
where $p$ denotes the order of autoregression, $q$ denotes the order of moving average and $\lbrace\epsilon_{t}\rbrace$ denotes the noise sequence of iid random variables with $S\alpha S$ distribution with (cf) given in (\ref{2par}) for $\alpha \in (1,2]$ and $\sigma > 0$. Further, in the unit disc $\lbrace z:|z|\leq 1 \rbrace$, $\phi(z)=1-\phi_{1}z-\cdots -\phi_{p}z^{p}$ and $\theta(z)=1+\theta_{1}z+\cdots +\theta_{q}z^{q}$ have no common zeros. For more details on ARMA($p,q$) process with $S\alpha S$ noise, (see \cite{sam}, pp. 376-380).

\noindent
Let the time series $\lbrace X_{t}\rbrace$ be a \textbf{GARCH($p,q$) process with $S\alpha S$ noise} denoted by $S\alpha S$-GARCH($p,q$), if
\begin{equation}
 X_{t} = \sigma'_{t}\epsilon_{t}, ~\sigma'_{t}=c+\sum_{i=1}^{q}a_{i}|X_{t-i}|+\sum_{j=1}^{p}b_{j}\sigma'_{t-j} \label{20}
\end{equation}
where $c>0$, $\boldsymbol{a}$ and $\boldsymbol{b}$ are non-negative and $\lbrace\epsilon_{t}\rbrace$ denotes the noise sequence of iid random variables with $S\alpha S$ distribution with (cf) given in (\ref{2par}). It is known that the $S\alpha S$-GARCH($1,1$) process has a unique strictly stationary solution if $b_1+\lambda a_1 <1$ while $S\alpha S$-GARCH($p,q$) process with $p \geq 2$ or $q\geq 2$ has a unique strictly stationary solution if $c>0$ and $\lambda \sum_{i=1}^{q}a_i+\sum_{j=1}^{p}b_j \leq 1$. For further details and information on the approximate values of $\lambda$ used for different values of $\alpha$, see \cite{mit}.

\noindent
We say that the time series $\lbrace X_{t}\rbrace$ is an \textbf{ARMA($p_{A},q_{A}$) process with $S\alpha S$-GARCH($p_{G},q_{G}$)} noise, denoted by ARMA($p_{A},q_{A}$)-$S\alpha S$-GARCH($p_{G},q_{G}$), if
\begin{equation}
X_{t}=\sum_{i=1}^{p_{A}}\phi_{i}X_{t-i}+\sum_{j=1}^{q_{A}}\theta_{j} e_{t-j}+e_{t},~~~ \\
e_{t}=\sigma'_{t}\epsilon_{t},~~~\sigma'_{t}=c+\sum_{i=1}^{q_{G}}a_{i}|e_{t-i}|+\sum_{j=1}^{p_{G}}b_{j}\sigma'_{t-j}\label{30}
\end{equation}
where $c>0$, $\boldsymbol{a}_G$ and $\boldsymbol{b}_G$ are non-negative and $\lbrace\epsilon_{t}\rbrace$ denotes the noise sequence of iid random variables with $S\alpha S$ distribution with (cf) given in (\ref{2par}) for $\alpha \in (1,2]$ and $\sigma > 0$.

\noindent
Note now that due to undefined covariance when $\alpha < 2$, the classical autocovariance function cannot be considered as a tool for developing methods of parameter estimation of the processes defined in (\ref{10}) and (\ref{30}). Thus, autocovariation (normalized) and autocodifference serves as a \textbf{measure of dependence} best suited for heavy-tailed distributions and processes. In our proposed method of estimating the parameters of the processes defined in (\ref{10}) and (\ref{30}), we make use of the normalized autocovariation. The normalized autocovariation (see \cite{gb}) for $S\alpha S$ process $\lbrace X_{t}\rbrace$ with lag $k$ is given by
\begin{equation}
NCV(X_{t},X_{t-k})=\frac{\mathbb{E}(X_{t}\text{sign}(X_{t-k}))}{\mathbb{E}|X_{t-k}|}\label{50}
\end{equation}
and its estimator for a sample $x_{1},\cdots,x_{N}$ being a realization of a stationary process $\lbrace X_{t}\rbrace$ is given by
\begin{equation}
\widehat{NCV}(X_{t},X_{t-k})=\frac{\sum_{t=l}^{r}x_{t}\text{sign}(x_{t-k})}{\sum_{t=1}^{N}|x_{t}|}
\end{equation}
where $l=\text{max}(1,1+k)$ and $r=\text{min}(N,N+k)$. 
\section{Parameter Estimation }\label{sec3}
We propose two methods for the estimation of the parameters $\boldsymbol \phi,~ \boldsymbol \theta,~ c,~ \boldsymbol a $, $\boldsymbol b$, $\boldsymbol{a}_G $ and $\boldsymbol{b}_G$ for the processes defined in (\ref{10}), (\ref{20}) and (\ref{30}). For estimation of $\boldsymbol \phi$ and  $\boldsymbol \theta$, we suitably modify the Hannen-Rissanen method \cite{han} and for estimating $c,~\boldsymbol{a}$, $\boldsymbol{b}$, $\boldsymbol{a}_G $ and $\boldsymbol{b}_G$ we modify the empirical characteristic function method discussed in \cite{om}. 
\subsection{Modified Hannan-Rissanen Method (MHR)} \label{sec3.1}
Let $\lbrace X_{t}\rbrace$ be as defined in (\ref{10}) or (\ref{30}). We discuss the Modified Hannan-Rissanen (MHR) algorithm  in three steps.
\begin{enumerate}
\item[ \textbf{Step 1.}]  Fit a  high order AR($a$) model to $\lbrace X_{t}\rbrace$ (with $a>$max($p,q$)) using the modified Yule-Walker estimation method, for $S\alpha S$ autoregressive models with $\alpha \in (1,2]$, see \cite{yule}.
\begin{enumerate}
\item Let  $\lbrace X_{t}\rbrace$ be an AR($a$) process given by
\begin{equation}
X_{t}-\phi_{1}X_{t-1}-\cdots-\phi_{a}X_{t-a}=\epsilon_{t} \label{1}
\end{equation}
where $\lbrace \epsilon_{t} \rbrace$ constitutes sample of an iid $S\alpha S$ random variables with $\alpha >1$.
\item Multiply both sides of (\ref{1}) by the vector $\textbf{S}=[S_{t-1},\cdots,S_{t-a}]$ where $S_{t}=\text{sign}(X_{t})$ and take the expectation to  obtain $a$ equations of the form 
\begin{equation}
\mathbb{E}X_{t}S_{t-j}-\sum_{i=1}^{a}\phi_{i}\mathbb{E}X_{t-i}S_{t-j}=\mathbb{E}\epsilon_{t}=0 ,~j =1,\cdots,a\label{2}
\end{equation}
\item Divide the $j$th equation of (\ref{2}) by $\mathbb{E}|X_{t-j}|$ to obtain 
\begin{equation}
\frac{\mathbb{E}X_{t}S_{t-j}}{\mathbb{E}|X_{t-j}|}-\sum_{i=1}^{a}\phi_{i}\frac{\mathbb{E}X_{t-i}S_{t-j}}{\mathbb{E}|X_{t-j}|}=0 ,~j =1,\cdots,a\label{3}
\end{equation}
\item Applying normalized autocovariation as given in (\ref{50}) to (\ref{3}), we obtain the following matrix form
$$\boldsymbol \lambda=\Lambda \Phi$$
where $\boldsymbol \lambda$ and $\Phi$ are vectors of length $a$ defined as
$$\boldsymbol \lambda=[NCV(X_{t},X_{t-1}),\cdots,NCV(X_{t},X_{t-a})]',~\Phi=[\phi_{1},\cdots,\phi_{a}]'$$
and the matrix $\Lambda$ has size $a\times a$ with the $(i,j)$th elements given by 
$$\Lambda(i,j)=NCV(X_{t},X_{t-i+j}).$$
\item Finally, in order to estimate the values of the parameter vector $\Phi$, the matrix $\Lambda$ should be nonsingular and make use of the sample normalized autocovariation $\widehat{NCV}$. Thus,
$$\hat{\Phi}=\hat{\Lambda}^{-1}\hat{\boldsymbol\lambda}.$$
\end{enumerate}
\item[\textbf{Step 2.}]
Using the obtained estimated coefficients  $\hat{\phi}_{a1},\cdots,\hat{\phi}_{aa}$, compute the estimated residuals from 
$$\hat{\epsilon}_{t}=X_{t}-\hat{\phi}_{a1}X_{t-1}-\cdots-\hat{\phi}_{aa}X_{t-a}, ~~t=a+1,\cdots,n$$.
\item[\textbf{Step 3.}]
Next, we estimate the vector of parameters, $\boldsymbol{\beta} =(\boldsymbol\phi,\boldsymbol\theta)$ by least absolute deviation regression of $X_{t}$ onto $(X_{t-1},\cdots, X_{t-p},\hat{\epsilon}_{t-1},\cdots,\hat{\epsilon}_{t-q}),~t=a+1+q,\cdots,n$ by minimizing $S(\boldsymbol{\beta)}$ with respect to $\boldsymbol\beta$ where,
$$S(\boldsymbol{\beta)}=\sum_{t=a+1+q}^{n}|X_{t}-\phi_{1}X_{t-1}-\cdots-\phi_{p}X_{t-p}-\theta_{1}\hat{\epsilon}_{t-1}-\cdots-\theta_{q}
\hat{\epsilon}_{t-q}|.$$
\end{enumerate} 
\begin{remark}
The proposed MHR method is both computationally efficient and nearly optimal in estimating the ARMA coefficients of (\ref{10}) or (\ref{30}). It is important to note that the MHR method does not require information (or estimation) of the parameters of noise distribution which is also true for LAD method. The effectiveness of the proposed method in comparison to other methods is shown in Section \ref{sec4}.           
\end{remark}
\subsection{Modified Empirical Characteristic Function Method (MECF)}\label{sec3.2}
We employ this method to obtain the estimates of $\boldsymbol a$, $\boldsymbol b$ and $c$  for processes defined in (\ref{20}) and (\ref{30}). For simplicity, we shall discuss the method for the process defined in (\ref{20}), for the case $p=1$ and $q=1$ relevant in empirical finance. The method can be extended to higher orders of $p$ and $q$ in a similar spirit. For $p=1$ and $q=1$ we get 
\begin{equation}
X_{t}=\sigma'_{t}\epsilon_{t},~\sigma'_{t}=c+a_{1}| X_{t-1}|+b_{1}\sigma'_{t-1} \label{100}
\end{equation}
where $c>0$, $a_1$ and $b_1$ are non-negative satisfying the stationary condition $\lambda a_1 +b_1 <1$ and $\lbrace\epsilon_{t}\rbrace$ denotes the noise sequence of iid random variables with $S\alpha S$ distribution.
The parameter estimates for (\ref{100}) are obtained by minimizing the function $f$ over $c$, $a_1$ and $b_1$ defined as
$$f(c,a_1,b_1)=\sum_{j=1}^{n}|\psi_{theoretical}(\hat{\epsilon}_{j})-\psi_{empirical}(\hat{\epsilon}_{j})|
$$
where $\hat{\epsilon}_{j}=\frac{X_{j}}{c+a_{1}| X_{j-1}|+b_{1}\sigma'_{j-1}}$, $\psi_{theoretical}(\hat{\epsilon}_{j})=\exp(-|\hat{\epsilon}_{j}|^{\hat{\alpha}})$, $\psi_{empirical}(\hat{\epsilon}_{j})=\frac{1}{n}\sum_{j=1}^{n}\cos(\hat{\epsilon}_{j}Y_{j})$ where $Y_1,\cdots, Y_n$ are iid random variables with $S\hat{\alpha}S$ distribution and $\hat{\alpha}$ is the estimate of $\alpha$ obtained from the noise sequence of iid random variables with $S\alpha S$ distribution using the hybrid method discussed in \cite{hybrid}. We make use of the function ``optim" available in ``stats" package of R for the minimization of the function $f$ over $c$, $a_1$ and $b_1$. The method used is ``L-BFGS-B" introduced by  Byrd et al. \cite{fg}. 
\section{Simulation and Comparative Analysis}
\label{sec4}
For the comparative analysis, we investigate the performance of LAD, LS and MHR for the estimation of the parameters of the processes defined in (\ref{10}) and (\ref{30}). The parameters of the process defined in (\ref{20}) are obtained using MECF discussed in Section \ref{sec3.2}.
\noindent
We consider four models to check the efficiency of our proposed method for (\ref{10}) and (\ref{30}) in comparison to LAD and LS through Monte Carlo simulations.
\begin{itemize}
\item[(\textbf{M1})] $S\alpha S$-MA(1): $X_{t}=\theta_{1}\epsilon_{t-1}+\epsilon_{t}$, $|\theta_{1}|< 1$
\item[(\textbf{M2})]$S\alpha S$-ARMA(1,1): $X_{t}=\phi_{1}X_{t-1}+\epsilon_{t}+\theta_{1}\epsilon_{t-1}$, $|\phi_{1}|< 1$ and $|\theta_{1}|< 1$
\item[(\textbf{M3})] MA(1)-$S\alpha S$-GARCH(1,1): $X_{t}=\theta_{1}\epsilon'_{t-1}+\epsilon'_{t}$, $|\theta_{1}|< 1$, $\epsilon'_{t}=\sigma'_{t}\epsilon_{t}$ with $\sigma'_{t}=c+a_{1}|X_{t-1}|+b_{1}\sigma'_{t-1}$ such that $\lambda a_1+b_1<1$.
\item[(\textbf{M4})] ARMA(1,1)-$S\alpha S$-GARCH(1,1): $X_{t}=\phi_{1}X_{t-1}+\theta_{1}\epsilon'_{t-1}+\epsilon'_{t}$, $|\theta_{1}|< 1$, $|\phi_{1}|< 1$, $\epsilon'_{t}=\sigma'_{t}\epsilon_{t}$\\ with $\sigma'_{t}=c+a_{1}|X_{t-1}|+b_{1}\sigma'_{t-1}$ such that $\lambda a_1+b_1<1$.
\end{itemize}
The noise sequence $\lbrace\epsilon_{t}\rbrace$ is considered to be
 iid  $S\alpha S$, where $\alpha \in (1,2]$ and  $\beta$, $\sigma$ and $\delta$ are fixed to 0, 1, 0 respectively and is generated using the function ``rstable" in the ``stabledist" package of R. For a selected set of values $\theta_{1}$, $\phi_{1}$ and $\alpha$, simulations are run where 1000 realisations of each model of length 1000 are generated. Tables \ref{11} and \ref{12} show the efficiency and effectiveness of our proposed MHR method over LAD and LS for models (\textbf{M1}) and (\textbf{M2}) respectively for different values of $\alpha \in [1.5,2]$, $\theta_{1}\in (0,0.5]$ and $\phi_{1} \in (0,1)$ in terms of mean and root mean squared error (RMSE). The estimate of $\alpha \in [1.5,2]$ for model (\textbf{M1}) and (\textbf{M2}) is obtained from the estimated residuals given in Step 2 of Subsection \ref{sec3.1} using the hybrid method discussed in \cite{hybrid} .

\noindent
Tables \ref{13} and \ref{14} show the accuracy and efficiency of the proposed MHR method and the MECF used in obtaining the estimate of $\theta_1$ and $(c,a_1,b_1)$ respectively for model (\textbf{M3}). The estimate of $\alpha$ for this model is obtained from the  $S\alpha S$ noise sequence $\lbrace\epsilon_t\rbrace$ using the hybrid method.

\noindent
Tables \ref{15} and \ref{16} show the accuracy and efficiency of the proposed MHR method and the MECF used in obtaining the estimate of $(\theta_1,\phi_1)$ and $(c,a_1,b_1)$ respectively for model (\textbf{M4}). The hybrid method is employed in obtaining the estimate of $\alpha$ from the $S\alpha S$ noise sequence $\lbrace\epsilon_t\rbrace$.

\noindent
We observe that for all the four models (\textbf{M1}), (\textbf{M2}), (\textbf{M3}) and (\textbf{M4}), the least squares method (LS) performs the worst while the least absolute deviation (LAD) and the proposed modified Hannan-Rissanen method (MHR) are at par with each other in terms of the robustness, accuracy and efficiency of the estimates. The estimates of the GARCH parameters obtained via MECF are also precise and accurate. 
\section{Application to Financial Data}
\label{sec5}
\begin{figure}[H]
\center
\includegraphics[keepaspectratio=true, scale=0.79]{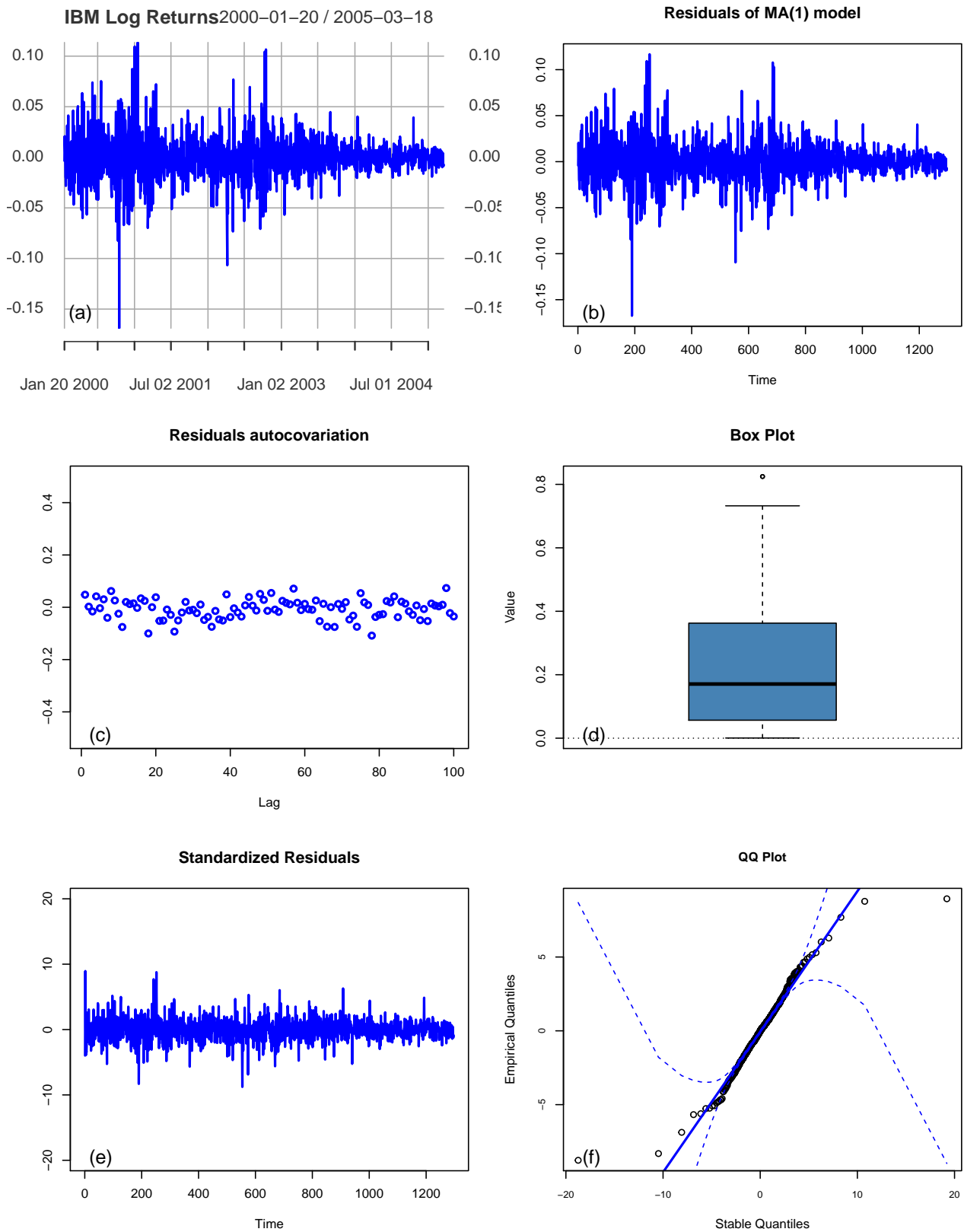}
\caption{ Plots generated using our proposed methods  }
\label{700}
\end{figure}

The time series real dataset that we have considered is the  International Business Machines Corporation (IBM) obtained from Yahoo Finance for the period January 19, 2000 - March 19, 2005, comprising of 1297 daily log returns values of the adjusted closing price. From Figure \ref{700} (a) , we observe the presence of time varying volatility of the log returns of the time series with mean almost zero and the minimum and maximum approximately around zero. The kurtosis is 5.8164 while the skewness is -0.030 which implies the distribution of the log returns is fairly symmetrical with heavier tails than the normal. These characteristics thereby suggest the application of ARMA($p_A,q_A$)-$S\alpha S$-GARCH($p_G,q_G$) model to the given dataset.
\noindent
To determine whether the considered time series data is stationary, we implement the Augmented Dickey-Fuller Test (ADF) ``adf.test" available in the ``tseries" package in R. The $p$-value obtained is 0.01 which confirms the stationarity of the data. We first try to implement the ARMA($p_A,q_A$) model and check whether it fits the dataset well. 
In order to select the best order for fitting an ARMA($p_A,q_A$) model  according to either AIC (Akaike Information Criterion) or BIC (Bayesian Information Criterion), we make use of ``auto.arima" available in the package ``forecast" in R. The best model chosen by ``auto.arima" was MA(1) model and we assume that the residuals obtained from this model constitute a sample from $S\alpha S$ distribution. Thus, we can employ the MHR method to obtain the estimate $\hat{\theta}_{1}=-0.078$.

\noindent
In order to check if the residuals can be considered as an independent sample, we make use of the empirical autocovariation function instead of the classical autocovariance function. From Figure \ref{700} (c), one can observe that the dependence in the residual series is almost unidentifiable.
\noindent
Finally, we prove that the distribution of the residual series is $S\alpha S$ using the Kolmogorov-Smirnov (KS) test and estimate the parameters ($\alpha$, $\beta$, $\sigma$ , $\delta$) using the  method in \cite{hybrid}. The estimates obtained for the residual series are ($\hat{\alpha}$, $\hat{\beta}$, $\hat{\sigma}$, $\hat{\delta}$)=$(1.6762,~0.0373,~0.0115,~0.0001)$. The KS test statistics is defined as:
$$D=\sup_{x}|G_{n}^{1}(x)-G_{n}^{2}(x)|,$$
where  $G_{n}^{1}(\cdot)$ and $G_{n}^{2}(\cdot)$ denote the empirical cumulative distribution function for Sample 1 and Sample 2 respectively, both of length $n$. The test derived from the KS statistics is called the two sample KS test.  In our case, Samples 1 and 2 are the residuals and the random sample simulated from $S\alpha S$ distribution with estimated parameters corresponding to the residuals respectively. Finally, we obtain 100 $p_{values}$ of the test and create a boxplot as shown in Figure \ref{700} (d). Assuming 0.05 confidence level, we fail to reject the hypothesis that the samples are from the same distribution.

\noindent
However, from the residuals plot of MA(1) with $S\alpha S$ noise, we observe that the volatility decays with time, thereby suggesting the replacement of $S\alpha S$ noise with $S\alpha S$-GARCH(1,1) noise in the $S\alpha S$-MA(1) model. Therefore, we will build MA(1)-$S\alpha S$-GARCH(1,1) model. For the given time series, the  initial values of $c$, $a_1$ and $b_1$ is considered to be 0.001, 0.1 and  0.8 respectively to obtain the estimates.  The GARCH(1,1) estimated parameters obtained after employing MECF are given in Table \ref{17}.

\noindent
In order to evaluate if the considered model is correctly specified, we analyse the estimated standardized residuals using graphical diagnostics. We plot the standardized residuals to check the  absence of conditional hetroskedascity. From Figure \ref{700} (e),  we observe that the standardized residual plot seems quite stable, indicating a constant variance over time.
Further,we make use of the QQ plot to check if the distribution of the standardized residuals matches the assumed noise distribution ($S\alpha S)$ used in the estimation. From Figure \ref{700} (f), we observe that the distribution of of the standardized residuals is almost ($S\alpha S)$ with parameters ($\hat{\alpha}$, $\hat{\beta}$, $\hat{\sigma}$, $\hat{\delta}$)=$(1.82398,~0.03870,~1.01298,~0.02132)$.

\section{Concluding Remarks}\label{sec6}
To conclude, we make the following observations in relation to our proposed methods.
\begin{itemize}
\item[1.] In this article, in order to estimate the ARMA coefficents of ARMA($p,q$) process with $S\alpha S$ noise as defined in (\ref{10}) or ARMA($p_{A},q_{A}$) process with $S\alpha S$-GARCH($p_{G},q_{G}$) as defined in (\ref{30}), we use the Modified Yule-Walker method \cite{yule} in the Hannan-Rissanen method and obtain the estimates of the parameters using LAD regression.
\item[2.] The M-estimators namely, the LAD and ML  perform well when the noise is heavy-tailed (stable). However, the LAD estimator is preferred over the ML  as it is computationally efficient, robust and does not require information (or estimation) of the parameters of the noise distribution.  The proposed MHR method also inherits the same advantages of LAD especially for $\alpha \in [1.5,2]$, $\theta_{1}\in (0,0.5]$ and $\phi_{1} \in (0,1)$. For details on the limitations of LS and ML estimator, see \cite{cad}.
\item[3.] For the estimation of the GARCH coefficients for GARCH($p,q$) process with $S\alpha S$ noise defined in (3) and  ARMA($p_{A},q_{A}$) process with $S\alpha S$-GARCH($p_{G},q_{G}$) defined in (4), we introduce and discuss MECF for the case $p=1$ and $q=1$ for simplicity. The method is computationally efficient, robust and can further be extended to higher orders of $p$ and $q$. 
\item[4.] Finally, we give an application of our proposed methods, where the noise distribution of the dataset is considered to be $S\alpha S$ and later $S\alpha S$-GARCH due to volatility clustering. From Table 7, we observe the estimates are statistically significant and Figure 1 shows efficient modelling of the financial data using our proposed methods through residual analysis.
\end{itemize}
\newpage
\newpage
\begin{landscape}
\begin{table}[tbph]
\centering 
\begin{tabular}{| *{5}{c |}} 
\hline
\textbf{True Values} & \textbf{LAD}& \textbf{LS}& \textbf{MHR} & $\hat{\alpha}_\textbf{residuals}$ \\
\hline
$\theta_{1}=0.01,~\alpha =1.50$& 0.0100 (0.0152) & 0.0096 (0.0283)  & 0.0098 (0.0153)& 1.5011 (0.0529)\\
\hline
$\theta_{1}=0.05,~\alpha =1.55$& 0.0496 (0.0175) & 0.0490 (0.0258)  & 0.0492 (0.0177) & 1.5553 (0.0532)\\
\hline
$\theta_{1}=0.10,~\alpha =1.65$& 0.0997 (0.0216) & 0.0991 (0.0269)  &0.0990 (0.0218) & 1.6552 (0.0524)\\
\hline
 $\theta_{1}=0.40,~\alpha =1.85$& 0.3996 (0.0310) & 0.3992 (0.0295)  & 0.3965 (0.0312) & 1.8523 (0.0438)\\
\hline
$\theta_{1}=0.50,~\alpha =1.95$& 0.4987 (0.0362) & 0.4990 (0.0306)  & 0.4911 (0.0372) & 1.9500 (0.0310)\\
\hline
\end{tabular}
\caption{Mean and RMSE (in parentheses) of 1000 estimates of $\theta_{1}$ and $\alpha$ for \textbf{M1} model for different values of $\alpha$}
\label{11}

\vskip 2ex

%
%
%
\begin{tabular}{| *{8}{c |}} 
\hline
\textbf{True Values}&  \multicolumn{2}{| c }{\textbf{LAD}}  & \multicolumn{2}{| c }{\textbf{LS}}&  \multicolumn{2}{| c | } {\textbf{MHR}} & $\hat{\alpha}_\textbf{residuals}$\\
\hline
$(\phi_1,\theta_1,\alpha)$ & \textbf{Mean} $(\hat{\phi}_1,\hat{\theta}_1)$ & \textbf{RMSE} $(\hat{\phi}_1,\hat{\theta}_1)$ & \textbf{Mean}$(\hat{\phi}_1,\hat{\theta}_1)$ & \textbf{RMSE} $(\hat{\phi}_1,\hat{\theta}_1)$ & \textbf{Mean} $(\hat{\phi}_1,\hat{\theta}_1)$ & \textbf{RMSE} $(\hat{\phi}_1,\hat{\theta}_1)$ & \\
\hline
$(0.50,0.01,1.55)$&(0.4993,0.0105)& (0.0285,0.0329)& (0.4957,0.0133)& (0.0474,0.0564)& (0.5004,0.0088)& (0.0275,0.0316) & 1.5545 (0.0564)\\
\hline
$(0.40,0.05,1.65)$&(0.3994,0.0503) &(0.0427,0.0474)& (0.3956,0.0534)& (0.0584,0.0662)& (0.4020,0.0466)& (0.0430,0.0470) & 1.6528 (0.0525)\\
\hline
$(0.80,0.10,1.75)$&(0.7983,0.1013)& (0.0172,0.0306)&(0.7961,0.1024) &(0.0206,0.0359)&(0.7988,0.0979)& (0.0182,0.0313)& 1.7462 (0.0553)\\
\hline
$(0.90,0.20,1.75)$&(0.8985,0.2011)& (0.0104,0.0274)&(0.8964,0.2019)& (0.0129,0.0318)&(0.8982,0.1951)& (0.0129,0.0315)& 1.7460 (0.0568)\\
\hline
$(0.20,0.40,1.80)$&(0.1989,0.4009)& (0.0455,0.0535)&(0.1972,0.4018)& (0.0484,0.0575)&(0.2035,0.3917)& (0.0663,0.0681)& 1.8008 (0.0469)\\
\hline
$(0.10,0.20,1.80)$&(0.0983,0.2016) &(0.0926,0.0968)&(0.0950,0.2041)& (0.0984,0.1036)&(0.1071,0.1919)& (0.1031,0.1036)& 1.8010 (0.0454)\\
\hline
$(0.30,0.10,1.85)$&(0.2982,0.1014)& (0.0731,0.0786)&(0.2954,0.1035)& (0.0716,0.0784)&(0.3009,0.0983)& (0.0756,0.0790)& 1.8487 (0.0414)\\
\hline
\end{tabular} 
\caption{Mean and RMSE both (in parentheses) of 1000 estimates of $\phi_{1}$, $\theta_{1}$ and $\alpha$ for (\textbf{M2}) model for different values of $\alpha$}
\label{12}

\vskip 2ex

%
\begin{tabular}{| *{4}{c |}} 
\hline
\textbf{True Values} & \textbf{LAD}& \textbf{LS}& \textbf{MHR} \\
\hline
$\theta_{1}=0.05,~\alpha =1.20$& 0.0508 (0.0247) & 0.0511 (0.0604)  &0.0464 (0.0225)\\
\hline
$\theta_{1}=0.20,~\alpha =1.40$& 0.1975 (0.0349) & 0.1961 (0.0653)  & 0.1925 (0.0354)\\
\hline
$\theta_{1}=0.30,~\alpha =1.65$& 0.3003 (0.0384) & 0.3014 (0.0577)  & 0.2989 (0.0382)\\
\hline
$\theta_{1}=0.40,~\alpha =1.75$& 0.3952 (0.0424) & 0.3935 (0.0449)  & 0.3893 (0.0405)\\
\hline
$\theta_{1}=0.50,~\alpha =1.85$& 0.5018 (0.0373) & 0.5074 (0.0376)  & 0.4942 (0.0367)\\
\hline
\end{tabular}
\caption{Mean and RMSE (in parentheses) of $\theta_{1}$ for (\textbf{M3}) model for different values of $\alpha$}
\label{13}

\vskip 2ex

%
%
%
\begin{tabular}{| *{5}{c |}} 
\hline
\textbf{True Values ($\theta_1,c,a_1,b_1,\alpha)$} & $\hat{c}$&$\hat{a}_1$& $\hat{b}_1$&$\hat{\alpha}_\textbf{noise}$\\
\hline
$(0.05,0.01,0.02,0.7,1.20)$& 0.0085 (0.0037) & 0.0196 (0.0035)  &0.7086 (0.0275) & 1.2011 (0.0471)\\
\hline
$(0.20,0.05,0.04,0.9,1.40)$& 0.0499 (0.0032) & 0.0399 (0.0032)  & 0.9060 (0.0246) & 1.4017 (0.0511)\\
\hline
$(0.30,0.10,0.05,0.8,1.65)$& 0.0121 (0.0879) & 0.0498 (0.0032)  & 0.8046 (0.0234) & 1.6548 (0.0552)\\
\hline
$(0.40,0.50,0.06,0.8,1.75)$& 0.4966 (0.0335) & 0.0607 (0.0033)  & 0.8102 (0.0272) & 1.7584 (0.0496)\\
\hline
$(0.50,1.00,0.03,0.9,1.85)$& 1.0507 (0.3165) & 0.0301 (0.0033)  & 0.9115 (0.0267) & 1.8574 (0.0433)\\
\hline
\end{tabular}
\caption{Mean and RMSE (in parentheses) of $c$, $a_1$, $b_1$, $\alpha$ for (\textbf{M3}) model for different values of $\alpha$}
\label{14}
\end{table}

\end{landscape}

\begin{landscape}

\begin{table}[h]
\centering 
\begin{tabular}{| *{7}{c |}} 
\hline
\textbf{ True Values} &  \multicolumn{2}{| c }{\textbf {LAD}} &  \multicolumn{2}{| c }{\textbf {LS}} &\multicolumn{2}{| c |}{\textbf {MHR}}\\
\hline
ARMA(1,1)-GARCH(1,1)& \textbf{$\theta_{1}$} & \textbf{$\phi_{1}$} & \textbf{$\theta_{1}$}& \textbf{$\phi_{1}$}& \textbf{$\theta_{1}$}& \textbf{$\phi_{1}$} \\
\hline
 $\theta_{1}=0.10,~\phi_{1}=0.40,~\alpha =1.55$ &  0.1004 (0.0497)  & 0.4028 (0.0422)& 0.1013 (0.0850) & 0.4008 (0.0764) & 0.0873 (0.0476)& 0.4126 (0.0438) \\
 \hline
$\theta_{1}=0.10,~\phi_{1}=0.60,~\alpha =1.65$ &  0.0973 (0.0509)  & 0.5979 (0.0415)& 0.0917 (0.0763) & 0.5977 (0.0584) & 0.0921 (0.0510)& 0.6015 (0.0419) \\
 \hline
 $\theta_{1}=0.20,~\phi_{1}=0.50,~\alpha =1.70$ &  0.2007 (0.0568)  & 0.4944 (0.0417)& 0.1929 (0.0932) & 0.5025 (0.0651) & 0.1920 (0.0564)& 0.5007 (0.0456) \\
 \hline
$\theta_{1}=0.20,~\phi_{1}=0.90,~\alpha =1.75$ &  0.1957 (0.0505)  & 0.8971 (0.0155)& 0.1965 (0.0683) & 0.8964 (0.0192) & 0.1912 (0.0482)& 0.8964 (0.0172) \\
 \hline
 $\theta_{1}=0.30,~\phi_{1}=0.80,~\alpha =1.85$ &  0.2967 (0.0436)  & 0.8005 (0.0219)& 0.2995 (0.0420) & 0.7970 (0.0228) & 0.2935 (0.0428)& 0.7998 (0.0280) \\
 \hline
\end{tabular}
\caption{Mean and RMSE (in parentheses) of $\theta_{1}$ and $\phi_{1}$ for (\textbf{M4}) model for different values of $\alpha$  }
\label{15}
\end{table}
\begin{table}[h]
\centering 
\begin{tabular}{| *{5}{c |}} 
\hline
\textbf{True Values} ($\theta_1,\phi_1,c,a_1,b_1,\alpha)$ & $\hat{c}$&$\hat{a}_1$& $\hat{b}_1$&$\hat{\alpha}_\textbf{noise}$\\
\hline
$(0.1,0.4,0.01,0.02,0.7,1.55)$& 0.0115 (0.0042) & 0.0196 (0.0040)  &0.7096 (0.0305) & 1.5572 (0.0554)\\
\hline
$(0.1,0.6,0.05,0.04,0.9,1.65)$& 0.0573 (0.0035) & 0.0438 (0.0085)  & 0.9141 (0.0302) & 1.6504 (0.0573)\\
\hline
$(0.2,0.5,0.10,0.05,0.8,1.70)$& 0.1090 (0.0879) & 0.0500 (0.0032)  & 0.8132 (0.0234) & 1.7036 (0.0471)\\
\hline
$(0.2,0.9,0.50,0.06,0.8,1.75)$& 0.5016 (0.0308) & 0.0605 (0.0030)  & 0.8133 (0.0263) & 1.7542 (0.0495)\\
\hline
$(0.3,0.8,1.00,0.03,0.9,1.85)$& 1.0622 (0.3311) & 0.0301 (0.0032)  & 0.9142 (0.0282) & 1.8507 (0.0485)\\
\hline
\end{tabular}
\caption{Mean and RMSE (in parentheses) of $c,~a_1,~b_1,~\alpha$ for (\textbf{M4}) model for different values of $\alpha$}
\label{16}
\end{table}
\begin{table}[h]
\centering 
\begin{tabular}{| *{5}{c |}} 
\hline
 & \textbf{Estimate} & \textbf{Standard Error} & \textbf{$t$-value}& \textbf{Pr($>|t|)$}\\
\hline
$c$ & 0.00107 & 0.00024 & 4.435 & 9.19 e-6 \\
\hline
$a_1$&0.09285 & 0.00082 & 112.71 & 0\\
\hline
$b_1$&0.79286 & 0.00103 & 766.69 &0\\
\hline
\end{tabular}
\caption{Estimation of $\theta_{1}$ for A.3 model for different values of $\alpha$}
\label{17}
\end{table}

\end{landscape}

\end{document}